\documentclass[12pt]{article} 
\usepackage[latin1]{inputenc}
\usepackage[german,english]{babel}
\usepackage{amssymb}
\usepackage{subfigure}
\usepackage{epsfig}
\usepackage{epsf}
\usepackage{graphicx,psfrag}
\usepackage{longtable}
\usepackage{rotating}

\begin{document}

\textbf{\Large Superposition in nonlinear wave and evolution equations}
\vspace{1 cm}

\textbf{H. W. Schürmann\footnotemark[1], V. S. Serov\footnotemark[2], J. Nickel\footnotemark[1]}

\vspace{1 cm}



\footnotetext[1]{Department of Physics, University of Osnabr\"uck, D-49069 Osnabr\"uck, Germany, corresponding author: jnickel@uos.de.}
\footnotetext[2]{Department of Mathematical Sciences, University of Oulu, FIN-90014 Finland.}

\small{
Real and bounded elliptic solutions suitable for applying the Khare-Sukhatme superposition procedure are presented and used to generate superposition
solutions of the generalized modified Kadomtsev-Petviashvili equation (gmKPE) and the nonlinear
cubic-quintic Schrödinger equation (NLCQSE).}\\

\textbf{KEY WORDS:} Linear superposition; solitary wave solution.\\
\textbf{PACS:} 02.30.Jr; 02.30.Gp

\section{Introduction}\label{Intro}

As has been shown recently \cite{38}, \cite{37}, \cite{37a}, \cite{37b}
 (periodic) Jacobian elliptic functions (if they are
solutions of a certain nonlinear wave and evolution equation (NLWEE)) are start solutions for
generating new solutions of the NLWEE by a linear superposition
procedure. Thus, elliptic functions are of specific importance for
finding solutions of NLWEEs. On the other hand, based on a symmetry reduction method, a technique to obtain elliptic
solutions of certain NLWEEs was proposed and applied to the gmKPE and the NLCQSE \cite{8a}, \cite{8b}, \cite{39}.

It is the aim of the present paper to combine these approaches in order to obtain general elliptic solutions that
can serve as start solutions for superposition ("suitable solutions").

The superposition procedure can be described as follows \cite{38}:
If a solution of a NLWEE$[\psi(x,y,t)] = 0$ can be expressed in terms of
Jacobian elliptic functions

\begin{equation}\label{Khare.f1}
\Psi(x,y,t) = \sum_{\nu = 0}^l a_{\nu} \textrm{qn}^{\nu}\left[\mu (x + k y + v t),
m\right],
\end{equation}

 where $\textrm{qn}$ is anyone of the Jacobian elliptic functions and
 $a_{\nu}, \mu, c$ are constants, then the superposition solution \cite[Eqs. (4), (14)]{38}

\begin{equation}\label{Khare.f2}
 \widetilde{\Psi}(x,t) = \sum_{\lambda = 1}^p \sum_{\nu = 0}^l a_{\nu} \textrm{qn}^{\nu}\left[\mu (x + k y + v_{p} t) + \frac{n (\lambda - 1) K(m)}{p}, m\right],
\end{equation}

 where $n \in \{2, 4\}$ (depending on the periodicity of the Jacobian elliptic function and on $\nu$)
 and $K(m), m$ denote the complete elliptic integral of first
 kind and the modulus parameter $(0 \leq m \leq1)$, respectively,
 also may be a solution of the NLWEE. The number $p$ is integer (it depends on the NLWEE whether it is even or/ and
 odd) and the speed $v_{p}$ can be determined by using certain
 remarkable, recently established, identities involving Jacobian
 elliptic functions \cite{37}, \cite{37a}. It should be noted,
 that the existence of solutions (\ref{Khare.f1}) of a certain
 NLWEE does not necessarily imply the existence of a solution
 (\ref{Khare.f2}). As shown in Refs. \cite{38}, \cite{37b}
solutions (\ref{Khare.f2}) exist for the Korteweg-de Vries
equation (KdV), the Kadomtsev-Petviashvili equation (KP), the
nonlinear (cubic) Schrödinger equation (NLSE), the $\lambda
\phi^4-$field equation, the Sine-Gordon equation and the
Boussinesq equation. On the other hand, it may happen, as will be
seen below, that a solution (\ref{Khare.f1}) is known but does not
lead to a solution (\ref{Khare.f2}). It is crucial for the
procedure, that appropriate relations between Jacobian elliptic
functions are known.

\vspace{1 cm}

The symmetry reduction approach can be outlined as follows
\cite{39}:

The NLWEE$[\psi(x,t)] = 0$ is reduced by an appropriate
transformation $\psi \rightarrow f$ (e.~g., $\psi(x,t) = f(z), z =
x - c t$), where $f$ is supposed to obey the ordinary nonlinear differential
equation ("basic equation")

\begin{equation}\label{Out.f1}
\left( \frac{ df(z)}{dz}\right)^2 = \alpha f^4 + 4\beta f^3+6\gamma
f^2+4\delta f+\epsilon\equiv R(f),
\end{equation}

(with real $z$, $f(z)$, $\alpha$, $\beta$, $\gamma$, $\delta$,
$\epsilon$), leading to an equation $P(f) =
0$, where $P$ denotes a polynomial in $f$. Vanishing coefficients
in the polynomial equation $P(f) = 0$ imply equations which
partly determine the coefficients $\alpha$, $\beta$, $\gamma$,
$\delta$, $\epsilon$ in Eq. (\ref{Out.f1}). In general, the
coefficients depend on the structure and parameters of the NLWEE
and, finally, on the parameters of the transformation $\psi \rightarrow f$. Thus, the problem of finding a solution of the NLWEE is reduced to finding
an appropriate transformation that leads to the basic equation (\ref{Out.f1}).\\

As is well known \cite{10} the solution $f(z)$ of Eq. (\ref{Out.f1}) can be written as

\begin{equation}\label{Out.f7}
f(z)=f_0+ \frac { R'(f_0) } {4[\wp(z;g_2,g_3)-\frac{1}{24}
R''(f_0)]}, \quad
\end{equation}

where $f_0$ is a simple root of $R(f)$ \cite{genSol} and the prime denotes differentiation with respect to $f$.

The invariants $g_2,g_3$ of Weierstrass' elliptic function
$\wp(z;g_2,g_3)$ are related to the coefficients of $R(f)$ by
\cite{11}

\begin{equation}\label{Out.f5}
g_2= \alpha\epsilon-4\beta\delta+3\gamma^2, \quad
\end{equation}

\begin{equation}\label{Out.f6}
g_3= \alpha\gamma\epsilon+2\beta\gamma\delta-\alpha\delta^2
-\gamma^3-\epsilon\beta^2. \quad
\end{equation}

The discriminant (of $\wp$ and $R$ \cite{11})

\begin{equation}\label{Out.f8}
\triangle= g_2^3- 27 g_3^2, \quad
\end{equation}

is suitable to classify the behaviour of $f(z)$. The conditions

\begin{equation}\label{Out.f16}
\triangle\neq 0\quad {\mathrm {or}}\quad \triangle=0, \quad
g_2>0,\quad g_3>0.
\end{equation}

lead to periodic solutions \cite{39}, whereas the conditions
\cite{9}

\begin{equation}\label{Out.f9}
\triangle=0,\quad g_2\geq 0,\quad g_3\leq 0
\end{equation}

are associated with solitary wave like solutions.

Physical solutions $f(z)$ must be real and bounded. Considering
the phase diagram of $R(f)$ \cite{8a}, \cite{6a} one obtains conditions,
expressed in terms of the coefficients of the basic equation, that
determine physical solutions.
For convenience these conditions are referred to as the phase diagram conditions (PDC) in the following.

\section{Elliptic start solutions for superposition}\label{StartSol}

To apply the superposition procedure it is important to know whether a solution of the NLWEE according to
(\ref{Khare.f1}) exists. To check this it is useful to rewrite Weierstrass'
 function $\wp$ as \cite{9a}

\begin{equation}\label{Khare.f3}
 \wp(z) = e_{3} + \frac{e_{1} - e_{3}}{\textrm{sn}^2(\sqrt{e_{1} - e_{3}}z, m)},
\end{equation}

where $e_{1} \geq e_{2}
\geq e_{3}$  denote the roots of the equation

\begin{equation}\label{Khare.f4}
4 s^3 - g_{2} s - g_{3} = 0.
\end{equation}

 Substitution of Eq. (\ref{Khare.f3}) into Eq. (\ref{Out.f7})
 yields \cite{9b}

\begin{equation}\label{Khare.f5}
f(z) = \frac{(\alpha {f_{0}}^3 + 4 \beta {f_{0}}^2 + 2 e_{3} f_{0}
+ 5 \gamma f_{0} + 2 \delta)
 \textrm{sn}^2(\sqrt{e_{1} - e_{3}} z, m) + 2 (e_{1} - e_{3}) f_{0}}{(-\alpha {f_{0}}^2 - 2 \beta f_{0} + 2 e_{3} - \gamma)
  \textrm{sn}^2(\sqrt{e_{1} - e_{3}} z, m) + 2 (e_{1} - e_{3})},
\end{equation}

 with $m = \frac{e_2 - e_{3}}{e_{1} - e_{3}}.$ Comparison with
 Eq. (\ref{Khare.f1}) shows that

\begin{equation}\label{Khare.f6}
-\alpha {f_{0}}^2 - 2 \beta f_{0} + 2 e_{3} - \gamma = 0
\end{equation}

 is a necessary and sufficient condition that defines the subset
 of solutions (\ref{Khare.f1}).
 If $\alpha = 0$ holds the simple root $f_0$ of $R(f)$ can be
 choosen such that Eq. (\ref{Khare.f6}) and PDC are satisfied \cite{9aa}. If $\alpha \neq
 0$ and $\beta = \delta = 0$ Eq. (\ref{Khare.f6}) is
 satisfied also. If $\alpha \neq 0$ and $\beta \neq 0$, $\delta = \epsilon =
 0$, Eq. (\ref{Khare.f6}), PDC, and the condition $\triangle =
 0$, $g_3 > 0$ are not consistent, so that
 trigonometric functions (which are possible for $\triangle = 0$, $g_3 >
 0$) are not suitable for superposition, because $f(z)$ is a
 constant according to the general solution of Eq. (\ref{Out.f1}) \cite{genSol}.\\
 Equation (\ref{Khare.f6}) represents a relation between the parameters $\{\alpha, \beta, \gamma, \delta,
 \epsilon\}$ and thus determines a subset of parameters of the
 problem modelled by the NLWEE for which further solutions can be
 generated by superposition according to Eq. (\ref{Khare.f2}). Combining
 Eqs. (\ref{Khare.f5}) and (\ref{Khare.f6}) (with $\alpha = 0$) we obtain

\begin{equation}\label{Khare.f7}
 f(z) = \frac{2 e_{3} - \gamma}{2 \beta} + \frac{12 e_{3}^2 - 3 \gamma^2 + 4 \beta \delta}{4 \beta (e_{1} - e_{3})}
 \textrm{sn}^2(\sqrt{e_{1} - e_{3}} z, m),
\end{equation}

 where $e_1$, $e_{3}$ must be chosen as the largest and smallest roots of Eq.
 (\ref{Khare.f4}), respectively,
 so that the condition (\ref{Khare.f6}) is valid for a simple root $f_{0}$ of Eq. (\ref{Out.f1}) that satisfies the
 PDC.\\

Equation (\ref{Khare.f7}) can be evaluated explicitely subject to the two cases $\alpha = 0$ and $\alpha \neq 0$, $\beta = \delta = 0$,
respectively. If $\alpha = 0$ and, for simplicity, $\epsilon = 0$ the start solutions for superposition are

\begin{equation}\label{Simple.f4}
f(z) = \left\{ \begin{array}{ll} -\frac{3 \gamma + \sqrt{9
\gamma^2 - 16 \beta \delta}}{4 \beta} \textrm{dn}^2\left(\frac{1}{2}
\sqrt{3 \gamma + \sqrt{9 \gamma^2 - 16 \beta \delta}} z, \frac{2
\sqrt{9 \gamma^2 - 16 \beta \delta}}{3 \gamma + \sqrt{9 \gamma^2 -
16 \beta \delta}}\right),&
\beta \delta > 0,\:\gamma > 0,\\
   &  \\
\frac{4 \delta}{- 3 \gamma + \sqrt{9 \gamma^2 - 16 \beta \delta}}
\textrm{sn}^2\left(\frac{1}{2} \sqrt{- 3 \gamma + \sqrt{9 \gamma^2 - 16
\beta \delta}}z, \frac{3 \gamma + \sqrt{9 \gamma^2 - 16 \beta
\delta}}{3 \gamma - \sqrt{9 \gamma^2 - 16 \beta \delta}}\right), &
 \beta \delta > 0, \: \gamma < 0,\\
  & \\
-\frac{3 \gamma + \sqrt{9 \gamma^2 - 16 \beta \delta}}{4 \beta}
\textrm{cn}^2\left(\frac{(9 \gamma^2 - 16 \beta
\delta)^{\frac{1}{4}}}{\sqrt{2}} z, \frac{3 \gamma + \sqrt{9
\gamma^2 - 16 \beta \delta}}{2 \sqrt{9 \gamma^2 - 16 \beta
\delta}}\right), & \beta \delta < 0,
\end{array} \right.
\end{equation}

where the various possibilities to satisfy (\ref{Khare.f4}) and (\ref{Khare.f6}) have been taken into account and
$ \triangle = 4 \beta^2 \delta^2 (9 \gamma^2- 16 \beta \delta) > 0$ is necessary and sufficient to fulfill PDC \cite{39a}.

\vspace{1 cm}

If $\alpha \neq 0$, $\beta = \delta = 0$ the start solutions read

\begin{equation}\label{Simple.f10}
h(z) = \left\{ \begin{array}{ll} - \frac{3 \gamma + \sqrt{9
\gamma^2 - \alpha \epsilon}}{\alpha} \textrm{dn}^2\left(\sqrt{3 \gamma +
\sqrt{9 \gamma^2 - \alpha \epsilon}} z, \frac{2 \sqrt{9 \gamma^2 -
\alpha \epsilon}}{3 \gamma + \sqrt{9 \gamma^2 - \alpha
\epsilon}}\right),& \alpha < 0, \: \gamma > 0, \:  \epsilon < 0,\\
  & \\
\frac{\epsilon}{-3 \gamma + \sqrt{9 \gamma^2 - \alpha \epsilon}}
\textrm{sn}^2\left(\sqrt{- 3 \gamma + \sqrt{9 \gamma^2 - \alpha \epsilon}}
z, \frac{3 \gamma + \sqrt{9 \gamma^2 - \alpha \epsilon}}{3 \gamma
- \sqrt{9 \gamma^2 - \alpha \epsilon}}\right),& \alpha > 0, \:
\gamma < 0, \:
 \epsilon > 0,\\
    &  \\
-\frac{3 \gamma + \sqrt{9 \gamma^2 - \alpha \epsilon}}{\alpha}
\textrm{cn}^2\left(\sqrt{2} (9 \gamma^2 - \alpha \epsilon)^{\frac{1}{4}} z,
\frac{3 \gamma + \sqrt{9 \gamma^2 - \alpha \epsilon}}{2 \sqrt{9
\gamma^2 - \alpha \epsilon}}\right), &
 \alpha < 0,\:  \epsilon > 0,
\end{array} \right.
\end{equation}

where $ \triangle = 64 \alpha^2
\epsilon^2 (9 \gamma^2 - \alpha \epsilon) > 0$
and - according to the Cartesian sign rule - three numbers of sign reversals in
the sequence of coefficients of $R(h)$ or $\triangle > 0$ and
$\alpha > 0$ and two sign reversals to fulfill PDC.

To sum up, Eqs. (\ref{Simple.f4}) and (\ref{Simple.f10})
represent all elliptic solutions with $\alpha = 0$, $\epsilon = 0$ and
$\alpha \neq 0$, $\beta = \delta = 0$, respectively, that are
suitable for the the procedure suggested by
Khare and Sukhatme. "All elliptic" means that the solutions presented in
Refs. \cite{38}, \cite{37b} are particular cases of Eqs.
(\ref{Simple.f4}), (\ref{Simple.f10}). "Suitable" includes that the
superposition procedure may fail if solutions according to
Eq. (\ref{Simple.f4}) or (\ref{Simple.f10}) are inserted into
the NLWEE in question leading
to conditions that cannot be evaluated with respect to $v_p$ (cf.
Eq. (\ref{Khare.f2})) because the associated relations
between Jacobian functions are unknown (cf. Sec. \ref{gmKPE}). Examples to obtain superposition solutions
are presented in the following. Equation (\ref{Khare.f7})
can be evaluated in the same manner subject to the PDC to obtain
physical elliptic solutions if the simplifying assumption $\epsilon = 0$ does not hold.

\section{Superposition solutions of the generalized modified Kadomtsev-Petviashvili equation}\label{gmKPE}

The approach outlined in the previous section can be elucidated by investigation of the gmKPE (Superposition solutions of the NLCQSE are presented in \ref{AppA}.)

\begin{equation}\label{KP.f1}
 \psi_{xt} + \left(\left(a + b \psi^q\right) \psi^q
 \psi_x\right)_x + c \psi_{xxxx} - \sigma^2 \psi_{yy} = 0,
\end{equation}

where $a$, $b$, $c$, $q$, $\sigma^2$ are real constants. As shown previously \cite{8b} elliptic traveling-wave solutions
to Eq. (\ref{KP.f1}) exist. The set of these solutions is determined by

\begin{eqnarray}\label{KP.f2}
 \psi(x, y, t) &=& f(z)^{1/q}, q \neq 0, \nonumber\\
  z &=& x + k y + v t,\\
 f_z^2 &=& \alpha f^4 + 4 \beta f^3 + 6 \gamma f^2 + 4 \delta f + \epsilon,\nonumber
\end{eqnarray}

where $\alpha$, $\beta$, $\gamma$, $\delta$, $\epsilon$ are given by Eqs. (16a)-(16g) in Ref. \cite{8b}.
As shown above, the conditions $\alpha = 0$ or $\beta = \delta = 0$, $\alpha \neq 0$ lead to suitable solutions.
Imposing additionally the PDC and condition (\ref{Khare.f6}), respectively, the parameters of solutions (\ref{KP.f2}) are

\begin{eqnarray}
q = \frac{1}{2}, \alpha = -\frac{b}{12 c}, \beta = 0, \gamma =
\frac{k^2 \sigma^2 - v}{24 c}, \delta = 0, \epsilon \neq 0, c \neq
0,\label{KP.f3}\\
q = 1, \alpha = 0, \beta = -\frac{a}{12 c}, \gamma = \frac{k^2
\sigma^2 - v}{6 c},\epsilon = 0, c \neq 0,\label{KP.f4}\\
q = 1, \alpha = -\frac{b}{6 c}, \beta = 0, \gamma = \frac{k^2
\sigma^2 - v}{6 c}, \delta = 0, \epsilon \neq 0, c \neq 0,\label{KP.f5}\\
q = 2, \alpha = 0, \beta = -\frac{a}{6 c}, \gamma = \frac{2 (k^2
\sigma^2 - v)}{3 c}, \epsilon = 0, c \neq 0.\label{KP.f6}.
\end{eqnarray}

Referring to (\ref{KP.f3}) and (\ref{KP.f4}) first, solutions according to (\ref{Simple.f10}) and (\ref{Simple.f4}), respectively,
have to be evaluated. Inserting (\ref{KP.f3}) into (\ref{Simple.f10}), one obtains the
suitable start solutions

\begin{equation}\label{KP.f6a}
\psi(x, y, t) = \left\{ \begin{array}{ll} B_{1}\: \textrm{dn}^2\left[\mu_{1} (x + k y + v t), m_{1}\right],
\frac{b}{c} > 0, \: \frac{k^2 \sigma^2 - v}{c} > 0,\: \epsilon < 0,\\
 \\
B_{2}\: \textrm{sn}^2\left[\mu_{2} (x + k y + v t), m_{2}\right],
\frac{b}{c} < 0, \: \frac{k^2 \sigma^2 - v}{c} < 0,\: \epsilon > 0,\\
  \\
B_{3}\: \textrm{cn}^2\left[\mu_{3} (x + k y + v t), m_{3}\right],
\frac{b}{c} > 0,\: \epsilon > 0,
\end{array} \right.
\end{equation}

where $B_{j}$, $\mu_{j}$, $m_{j}$ are determined by inserting the parameters
$\alpha$, $\gamma$, $\epsilon$  according to (\ref{KP.f3}) into (\ref{Simple.f10}).
Formally the same result is obtained by inserting (\ref{KP.f4}) into (\ref{Simple.f4}).

Referring, secondly, to (\ref{KP.f5}) the solutions follow from (\ref{Simple.f10}) as

\begin{equation}\label{KP.f6c}
\psi(x, y, t) = \left\{ \begin{array}{ll} B_{1}\: \textrm{dn}\left[\mu_{1} (x + k y + v t), m_{1}\right],
\frac{b}{c} > 0, \: \frac{k^2 \sigma^2 - v}{c} > 0,\: \epsilon < 0,\\
 \\
B_{2}\: \textrm{sn}\left[\mu_{2} (x + k y + v t), m_{2}\right],
\frac{b}{c} < 0, \: \frac{k^2 \sigma^2 - v}{c} < 0,\: \epsilon > 0,\\
  \\
B_{3}\: \textrm{cn}\left[\mu_{3} (x + k y + v t), m_{3}\right],
 \frac{b}{c} > 0,\: \epsilon > 0,
\end{array} \right.
\end{equation}

where (again) $B_{j}$, $\mu_{j}$, $m_{j}$ are determined from (\ref{Simple.f10})
with parameters according to (\ref{KP.f5}). Formally the same results are given by (\ref{KP.f6}) and (\ref{Simple.f4}).

According to Eq. (\ref{Khare.f2}) the first solution in (\ref{KP.f6c}) leads to a superposition solution for $p = 2$

\begin{eqnarray}\label{KP.f13}
\widetilde{\psi}(x,y,t) &=& B \sum_{i = 1}^2
\textrm{dn}\left(\mu (x + k y + v_2 t) +  (i - 1) K(m),
m\right),\nonumber\\
B &=& B_{1},  \mu = \mu_{1}, m = m_{1}.
\end{eqnarray}

Inserting $\widetilde{\psi}(x, y, t)$ (denoting $\textrm{d}_i = \textrm{dn}\left(\mu (x + k y + v_2 t) +  (i - 1) K(m), m \right)$) into Eq. (\ref{KP.f1})
($a = 0$, because $\beta = 0$ according to (\ref{KP.f5})) we get

\begin{eqnarray}\label{KP.f14}
\left(- B m \mu v_2 - B c \mu^3 (2 m - m^2)\right) \frac{d}{dx}
\sum_{i = 1}^2 \textrm{s}_i \textrm{c}_i + \sigma^2 B m \mu k \frac{d}{d y} \sum_{i =
1}^2 \textrm{s}_i \textrm{c}_i\nonumber\\
+ 2 b B^3 m^2 \mu^2 \sum_{i = 1}^2 \textrm{d}_i \left(\sum_{i = 1}^2 \textrm{s}_i
\textrm{c}_i\right)^2 - m \mu b B^3 \left(\sum_{i = 1}^2 \textrm{d}_i\right)^2
\frac{d}{dx} \sum_{i = 1}^2 \textrm{s}_i \textrm{c}_i \\
+ 6 B c m \mu^3 \frac{d}{dx} \sum_{i = 1}^2 \textrm{d}_i^2 \textrm{s}_i \textrm{c}_i =
0.\nonumber
\end{eqnarray}

The last three terms of Eq. (\ref{KP.f14}) can be simplified as follows.

Using $\textrm{d}_1 \textrm{d}_2 = \sqrt{1 - m}$ and $\textrm{c}_1 \textrm{s}_1 \textrm{d}_2 +
\textrm{c}_2 \textrm{s}_2 \textrm{d}_1 = 0$ 
\cite[Eqs. (31), (39)]{37} we obtain

\begin{equation}\label{KP.f16}
 \left(\sum_{i = 1}^2 \textrm{d}_i\right)^2 \sum_{i = 1}^2 \textrm{s}_i \textrm{c}_i = \sum_{i =
 1}^2 \textrm{d}_i^2 \textrm{s}_i \textrm{c}_i + \sqrt{1 - m} \sum_{i = 1}^2 \textrm{s}_i \textrm{c}_i.
\end{equation}

Evaluating $\frac{d}{dx} \left(\left(\sum_{i = 1}^2 \textrm{d}_i\right)^2
\sum_{i = 1}^2 \textrm{s}_i \textrm{c}_i \right)$ and using Eq. (\ref{KP.f16}),
Eq. (\ref{KP.f14}) can be rewritten as

\begin{eqnarray}\label{KP.f17}
\left(- B m \mu v_2 - B c \mu^3 (2 m - m^2) - m \mu b B^3
\sqrt{1 - m}\right) \frac{d}{dx} \sum_{i = 1}^2
\textrm{s}_i \textrm{c}_i\\
+ \sigma^2 B m \mu k \frac{d}{dy} \sum_{i = 1}^2 \textrm{s}_i \textrm{c}_i + B m \mu \left(6 c \mu^2
-  b B^2\right)\frac{d}{dx} \sum_{i = 1}^2
\textrm{d}_i^2 \textrm{s}_i \textrm{c}_i = 0.\nonumber
\end{eqnarray}

The expression $(6 c \mu^2
-  b B^2)$ vanishes identically \cite{53}. With $\frac{d}{dy} \sum_{i = 1}^2 \textrm{s}_i \textrm{c}_i
= k \frac{d}{dx} \sum_{i = 1}^2 \textrm{s}_i \textrm{c}_i$ Eq. (\ref{KP.f17})
reads

\begin{equation}\label{KP.f18}
\left(-B m \mu v_2 - B c \mu^3 (2 m - m^2) - m \mu b B^3 \sqrt{1 -
m} + \sigma^2 B m \mu k^2\right) \frac{d}{dx} \sum_{i = 1}^2 \textrm{s}_i
\textrm{c}_i = 0,
\end{equation}

so that the speed $v_2$ is given by

\begin{equation}\label{KP.f19}
v_2 = - c \mu^2 (2 - m) - b B^2 \sqrt{1 - m} + \sigma^2 k^2.
\end{equation}

Thus, we have found a superposition solution of Eq.
(\ref{KP.f1}) for this particular case.

\begin{figure}[htbp]
\begin{center}
\includegraphics[width=.6\linewidth]{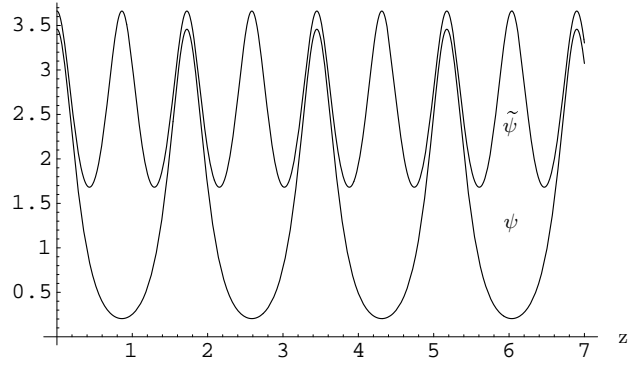}
\caption{\label{nick0404_Fig1}\small The start solution $\psi(z)$ (cf. first solution of Eq. (\ref{KP.f6c})) and the
superposition solution $\widetilde{\psi}(z)$ (cf. Eq. (\ref{KP.f13})) for
$\alpha = -2$, $\gamma = 4$, $\epsilon = -1$, $z = x + k y + v t$ and $z = x + k y + v_2 t$, respectively.}
\end{center}
\end{figure}

The start solution and the superposition solution are shown in Fig. \ref{nick0404_Fig1}.

We can generate superposition solutions for $p = 3$ from (\ref{KP.f6a}). As an example we consider the solution of the form
$\textrm{dn}^2$ in detail and compare it with the results of Cooper, Khare
and Sukhatme \cite{38}. According to Eq. (\ref{Khare.f2}) the
superposition ansatz reads

\begin{eqnarray}\label{KP.f24}
\widetilde{\psi}(x,y,t) &=& B \sum_{i = 1}^3 \textrm{dn}^2\left(\mu (x + k y + v_3 t) +  \frac{2 (i - 1) K(m)}{3}, m\right),\nonumber\\
B &=& B_{1}, \mu = \mu_{1}, m = m_{1}.
\end{eqnarray}

Inserting $\widetilde{\psi} (x,y,t)$ (denoting $\textrm{d}_i =
\textrm{dn}\left(\mu (x + k y + v_3 t) +  \frac{2 (i - 1)
K(m)}{3}, m\right)$) into Eq. (\ref{KP.f1}), we obtain

\begin{eqnarray}\label{KP.f25}
2 B m \mu (v_3 + 8 c \mu^2 - 4 c m \mu^2)\frac{d}{dx}
\sum_{i = 1}^3 \textrm{d}_i \textrm{s}_i \textrm{c}_i + 2 \sigma^2
B m \mu k \frac{d}{dy} \sum_{i =
1}^3 \textrm{d}_i \textrm{s}_i \textrm{c}_i\nonumber \\
+ 4 B^2 b m^2 \mu^2 \left(\sum_{i = 1}^3 \textrm{d}_i \textrm{s}_i
\textrm{c}_i\right)^2 - 2 B^2 b m \mu \sum_{i = 1}^3
\textrm{d}_i^2 \frac{d}{dx} \sum_{i = 1}^3 \textrm{d}_i
\textrm{s}_i \textrm{c}_i \\
+ 24 B c m \mu^3 \frac{d}{dx} \sum_{i = 1}^3 \textrm{d}_i^3
\textrm{s}_i \textrm{c}_i = 0.\nonumber
\end{eqnarray}

The last three terms can be rewritten as

\begin{equation}\label{KP.f25a}
- 2 B^2 b m \mu\left(-2 m \mu \left(\sum_{i = 1}^3 \textrm{d}_i
\textrm{s}_i \textrm{c}_i\right)^2 + \sum_{i = 1}^3 \textrm{d}_i^2
\frac{d}{dx} \sum_{i = 1}^3 \textrm{d}_i \textrm{s}_i
\textrm{c}_i - 12 \frac{c \mu^2}{b B} \frac{d}{dx} \sum_{i
= 1}^3 \textrm{d}_i^3 \textrm{s}_i \textrm{c}_i\right),
\end{equation}

whereas evaluation of $\frac{d}{dx} \left(\sum_{i = 1}^3
\textrm{d}_i^2 \sum_{j \neq i} \textrm{d}_j\textrm{s}_j \textrm{c}_j
\right)$ yields

\begin{eqnarray}\label{KP.f26}
\frac{d}{dx} \left(\sum_{i = 1}^3 \textrm{d}_i^2
\sum_{j \neq i} \textrm{d}_j \textrm{s}_j
 \textrm{c}_j\right)&=& - 2 m \mu \left(\sum_{i = 1}^3 \textrm{d}_i \textrm{s}_i \textrm{c}_i\right)^2 +
 \sum_{i = 1}^3 \textrm{d}_i^2 \frac{d}{dx} \sum_{i = 1}^3 \textrm{d}_i \textrm{s}_i \textrm{c}_i \nonumber \\
     & &+ 2 m \mu \sum_{i = 1}^3 \textrm{d}_i^2 \textrm{s}_i^2 \textrm{c}_i^2 - \sum_{i = 1}^3 \left(\textrm{d}_i^2 \frac{d}{dx} \textrm{d}_i \textrm{s}_i \textrm{c}_i\right)\\
           &=& - 2 m \mu \left(\sum_{i = 1}^3 \textrm{d}_i \textrm{s}_i \textrm{c}_i\right)^2 +
 \sum_{i = 1}^3 \textrm{d}_i^2 \frac{d}{dx} \sum_{i = 1}^3 \textrm{d}_i \textrm{s}_i \textrm{c}_i -
\frac{d}{dx} \sum_{i = 1}^3 \textrm{d}_i^3 \textrm{s}_i
\textrm{c}_i.\nonumber
\end{eqnarray}

Because $12 \frac{c \mu^2}{b B} = 1$ (in Eq. (\ref{KP.f25a})) holds identically,
we can use Eq. (\ref{KP.f26}) and \cite[Eq. (11)]{37b}

\begin{equation}\label{KP.f27}
\frac{d}{dx}\left(\sum_{i = 1}^3 \textrm{d}_i^2 \sum_{j
\neq i} \textrm{d}_j \textrm{s}_j \textrm{c}_j\right) = A(3, m)
\frac{d}{dx} \sum_{i = 1}^3 \textrm{d}_i \textrm{s}_i
\textrm{c}_i,
\end{equation}

to rewrite Eq. (\ref{KP.f25}) as

\begin{equation}\label{KP.f28}
-2 B m \mu (v_3 + 8 c \mu^2 - 4 c m \mu^2 + B b A(3,m))
\frac{d}{dx} \sum_{i = 1}^3 \textrm{d}_i \textrm{s}_i \textrm{c}_i\\
+ 2 \sigma^2 B m \mu k \frac{d}{dy} \sum_{i = 1}^3
\textrm{d}_i \textrm{s}_i \textrm{c}_i = 0.\nonumber
\end{equation}

Using $\frac{d}{dy} \sum_{i = 1}^3 \textrm{d}_i
\textrm{s}_i \textrm{c}_i = k \frac{d}{dx} \sum_{i = 1}^3
\textrm{d}_i \textrm{s}_i \textrm{c}_i$ this equation reads

\begin{equation}\label{KP.f29}
-2 B m \mu (v_3 + 8 c \mu^2 - 4 c m \mu^2 - \sigma^2 k^2 + B b
A(3,m)) \frac{d}{dx} \sum_{i = 1}^3 \textrm{d}_i
\textrm{s}_i \textrm{c}_i = 0.
\end{equation}

Thus, the speed $v_3$ in the superposition solution
(\ref{KP.f24}) (of a particular case) of Eq. (\ref{KP.f1}) is
given by

\begin{equation}\label{KP.f30}
 v_3 = 4 c m \mu^2 + \sigma^2 k^2 - 8 c \mu^2 - B b A(3, m).
\end{equation}

\begin{figure}[htbp]
\begin{center}
\includegraphics[width=.6\linewidth]{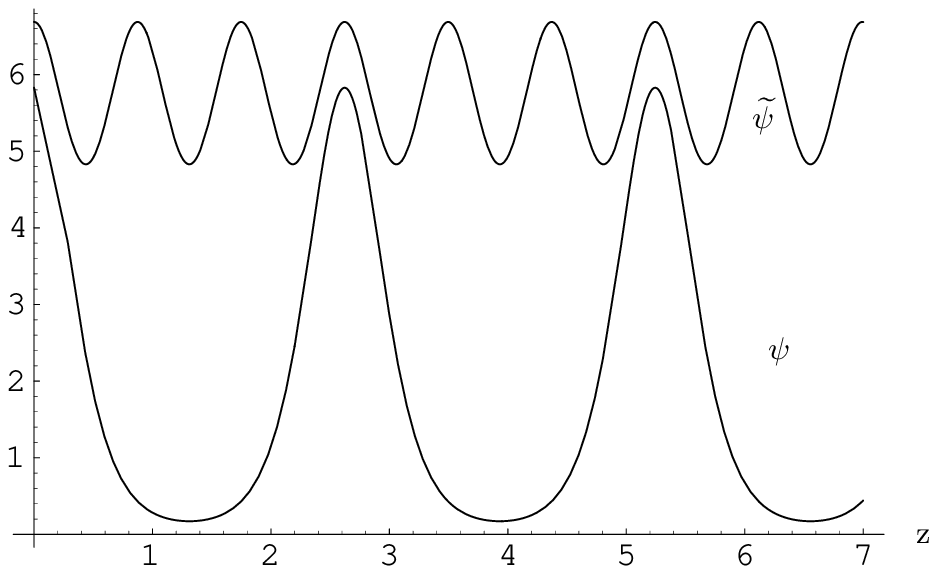}
\caption{\label{nick0404_Fig2}\small The start solution $\psi(z)$ (cf. first solution of Eq. (\ref{KP.f6a})) and the
superposition solution $\widetilde{\psi}(z)$ (cf. Eq. (\ref{KP.f24})) for
$\alpha = -1$, $\gamma = 1$, $\epsilon = -1$, $z = x + k y + v t$ and $z = x + k y + v_3 t$, respectively.}
\end{center}
\end{figure}

The start solution and the superposition solution are shown in Fig. \ref{nick0404_Fig2}.

Applying an analogous procedure with the ansatz

\begin{eqnarray}\label{KP.f31}
\widetilde{\psi}(x,y,t) &=& B \sum_{i = 1}^3 \textrm{sn}^2\left(\mu (x + k y + v_3 t) +  \frac{2 (i - 1) K(m)}{3}, m\right),\nonumber\\
B &=& B_{2}, \mu = \mu_{2}, m = m_{2}
\end{eqnarray}

and with the ansatz

\begin{eqnarray}\label{KP.f32}
\widetilde{\psi}(x,y,t) &=& B \sum_{i = 1}^3 \textrm{cn}^2\left(\mu (x + k y + v_3 t) +  \frac{2 (i - 1) K(m)}{3}, m\right),\nonumber\\
B &=& B_{3}, \mu = \mu_{3}, m = m_{3}
\end{eqnarray}

we obtain superposition solutions with

\begin{equation}\label{KP.f33}
 v_3 = 4 c m \mu^2 + 4 c \mu^2 + \sigma^2 k^2 + B b \frac{A(3, m) - 2}{m}
\end{equation}

for solution (\ref{KP.f31}) and

\begin{equation}\label{KP.f34}
v_3 = -8 c m \mu^2 + 4 c \mu^2 + \sigma^2 k^2 - B b \frac{A(3, m)
- 2 (1 - m)}{m}
\end{equation}

for solution (\ref{KP.f32}).

In deriving (\ref{KP.f33}) and (\ref{KP.f34}) we have used the
relations

\begin{equation}\label{KP.f35}
\frac{d}{dx}\left(\sum_{i = 1}^3 \textrm{s}_i^2 \sum_{j
\neq i}^3 \textrm{s}_j \textrm{c}_j \textrm{d}_j\right) =
-\frac{1}{m} \left(A(3,m) - 2\right) \frac{d}{dx} \sum_{i =
1}^3 \textrm{s}_i \textrm{c}_i \textrm{d}_i
\end{equation}

and

\begin{equation}\label{KP.f36}
\frac{d}{dx} \left(\sum_{i = 1}^3 \textrm{c}_i^2 \sum_{j
\neq i}^3 \textrm{s}_j \textrm{c}_j \textrm{d}_j\right) =
\frac{1}{m} \left(A(3,m) - 2 (1 - m)\right) \frac{d}{dx}
\sum_{i = 1}^3 \textrm{s}_i \textrm{c}_i \textrm{d}_i,
\end{equation}

respectively, which follow from Eq. (\ref{KP.f27}) and well known
relations between Jacobian elliptic functions.\\

Comparison of the Kadomtsev-Petivashvili equation together with the
ansatz considered by Cooper, Khare and Sukhatme \cite[Eqs. (1),(4)]{38}
with the Eqs. (\ref{KP.f1}), (\ref{KP.f3}) and (\ref{KP.f24}) shows that, apart from an additive constant \cite{37d}, our
result (\ref{KP.f30}) is consistent with that of Cooper, Khare and
Sukhatme \cite[Eq. (11), $\beta = 0$]{38} .
The cases related to (\ref{KP.f33}), (\ref{KP.f34}) have not been considered in Ref. \cite{38}.\\

To conclude, we note that real and bounded suitable solutions of the gmKPE only
exist for four different values of $q$ (cf. (\ref{KP.f3}) - (\ref{KP.f6})), though there is no
restriction for $q$ (apart from being real) of the known elliptic solutions of the gmKPE \cite{8b}.

The second of Eqs. (\ref{KP.f6c}) does not lead to a superposition solution although the solution has the form (\ref{Khare.f1}) \cite{Sug1}.
In this case, it seems that an appropriate identity for Jacobian elliptic functions does not exist. Thus, the claim at the end of Ref. \cite{38}
seems to strong.

\section{Summary and concluding remarks}

By combining the superposition principle and symmetry reduction we obtained general elliptic solutions suitable for superposition.
The results were applied to the gmKPE and the NLCQSE (see \ref{AppA}).
In Ref. \cite{38} particular elliptic solutions for
generating superposition solutions of the NLSE and the KPE were
used. As outlined above we start from (general) suitable solutions (cf. Eqs. (\ref{Simple.f4}), (\ref{Simple.f10}), (\ref{KP.f6a}),
(\ref{KP.f6c})) to obtain superposition solutions
more general than those in Ref. \cite{38}. We note that there are no
restrictions in advance for the coefficients of the NLSE and the
KPE. Constraints result from the condition that suitable solutions exist (cf.
Eq. (\ref{Khare.f6})) and from the PDC. As is obvious from the following list \ref{Table} there are rather many NLWEEs that exhibit
suitable elliptic solutions. Thus, it seems interesting to check whether they lead to superposition solutions by applying
Eqs. (\ref{Simple.f4}) and (\ref{Simple.f10}).


\renewcommand{\thetable}{\Roman{table}} 

\small{
\begin{longtable}{p{4 cm}| p{4 cm}| p{4.2 cm}|p{2 cm}} 
\caption{Elliptic solutions of various nonlinear wave and
evolution equations.} \label{Table}
\\
\hline
 Equation         &   Ansatz       &          Basic equation &      Suitable for\\
                       &                &                         &      superposition\\
 \hline
\endfirsthead
\hline
\endhead
\hline
\endfoot

\endlastfoot
$\psi_{t} -\psi\psi_{x}-\psi_{xxt}=0$& $\psi=f(kx-ct)=f(z)$ &
 $(f_z)^2= -\frac{f^3}{3kc} +\frac{f^2}{k^2}$&         \\

Benjamin-Bona-Mahony& & $+ 4\delta f +
\epsilon$& +\\\hline

$\psi_{tt} -\psi_{xx}+3(\psi^2)_{xx}$&
$\psi=f(kx-ct)=f(z)$  &
 $(f_z)^2= \frac{2f^3}{k^2} +\frac{c^2-k^2}{k^4}f^2$&        \\

$-\psi_{xxxx}=0$& & $ + 4\delta f + \epsilon$& +\\
Boussinesq& & &\\
\hline

  $\psi_t+\psi\psi_x-\psi_{xx}=0$ &  $\psi=f(kx-ct)=f(z)$   & $(f_z)^2=
  \frac{(2c^2f-kcf^2+4k^4\delta)^2}{4k^4c^2}$
  &    - \cite{Burgers}     \\
Burgers&&&\\\hline

$\psi_{tt} -\psi_{xx}+\sin\psi $&
$\psi=4\arctan [f(kx-ct)]$  &
 $(f_z)^2= 3 (\gamma+\frac{1}{8(c^2-k^2)})f^4 $&         \\

$+\frac{1}{2}\sin\frac{\psi}{2}=0$& $=4\arctan [f(z)]$& $+6\gamma f^2+3\gamma +\frac{5}{(c^2-k^2)},$& +\\

Double sine-Gordon& & $ c^2\neq k^2$&\\\hline

$\psi_{t}
+a\psi_{xx}-b\psi$& $\psi=$ & $(f_x)^2=  \frac{c_2}{3d\: a_1+a_2(2-d^2)}f^4$&          \\
$-c\mid\psi\mid^2\psi=0$&$f(x)\exp(ig(x))\exp(i\lambda t), $& & \\

$ a=a_1+ia_2, b=b_1+ib_2,$& $g_x(x)=d\frac{f_x}{f}$&
$+\left(\frac{\lambda_{\pm}-b_2}{d^2 a_2-a_2-2d\: a_1}\right)f^2$\cite{GinzLan}&+\\

$c=c_1+ic_2$&  &   &  \\

Ginzburg-Landau& & & \\\hline

$\psi_t + (b \psi +1) \psi_x + \psi_{ttx} = 0$& $\psi(x,t) = $& $(f_z)^2 = - \frac{b}{3 c^2} f^3 + \frac{c - k}{c^2 k} f^2$ &  \\
Joseph-Egri &$f(k x - c t) = f(z)$ & $- \frac{2 C_1}{c^2 k} f - \frac{2 C_2}{c^2 k}, $ & +\\
\cite{JoEg} & &$C_1$, $C_2$ const. &\\\hline

$\psi_t+ a\psi^2\psi_{x}+ b\psi_{x}\psi_{xx}+ $ & $\psi(x,t)=$ &  $(f_z)^2= 4\beta f^3 + 6\gamma f^2$& \\

$ g\psi\psi_{xxx} + \psi_{xxxx}=0$& $f(x-ct)=f(z)$& $+4\delta f+\epsilon$& +\\
  Korteweg-de Vries &  & &\\\hline

 $\psi_{t} -\psi_{xx}-6\psi\psi_{x}+\psi_{xxx}$&
$\psi=f^2(kx-ct)=f^2(z)$ &
 $(f_z)^2= \frac{f^2}{k^2}\cdot $&      \\

 $=0$& &$\cdot\left(\frac{f^2}{2}\pm
\frac{2f}{5\sqrt{2}}+\frac{1}{25}\right)$ &  -   \cite{KdVB}\\

Korteweg-de Vries-Burgers& & &\\\hline

 $\psi_t + a \psi \psi_x + \psi_{xxx}$& $\psi(x,y,z,t) = f(\xi)$& $(f_{\xi})^2 = - \frac{a}{3 p^2} f^3 $ & \\
$+ \psi_{xyy} + \psi_{xzz} = 0$&$\xi = k x + l y + m z + \omega
t$& $-
\frac{\omega}{k p^2} f^2 - \frac{2 C_1}{k p^2},$ & +\\
KdV-Zakharov-Kusnetzov  & & $p^2 = k^2 + l^2 + m^2$, & \\
\cite{ZakKus}& & $C_1$\ const.& \\ \hline

$V_t =\partial^3V + \overline{\partial}^3V $ & $V(x,y,t) = \psi(z)$ &
$(f_z)^2 = -\frac{8 a_2}{1 + k^2} f^3
$
&\\
$+ 3 \partial (u V) + 3 \overline{\partial} (\overline{u} V),$&$z = x + k y - v t$&$-\frac{24 a_0}{1 + k^2} f^2+
\frac{12 F}{3 (3 k^2 - 1)} f^2$ & +\\
$\overline{\partial} u = \partial V$ & &$  + 4 \delta f + \epsilon$,
&\\
Novikov-Veselov & & $F = v + 3 C_0 + 3 k C_1$;
&\\
& & $C_0$, $C_1$ const.&\\\hline

 $\psi_{xx} -\psi_{tt}-\sin\psi =0$&$\psi(x,t)= 4\arctan[\frac{X(x)}{T(t)}]$ & $(\frac{dX}{dx})^2=R_1(X)$& \\

  sine-Gordon&  &$ = \alpha X^4+6\gamma X^2+\epsilon$ & +\\

& & $(\frac{dT}{dt})^2=R_1(T)$& \\

& & $ = \alpha T^4+(6\gamma -1)T^2-\epsilon$&\\\hline

$i \psi_x + \psi_{tt} + 2 \sigma |\psi|^2 \psi $& $\psi(x,t) = f(z) e^{i (r x - \lambda t)}$ &
$(f_z)^2 = - \frac{\sigma}{c (c + k \mu)} f^4 $ & \\

$- \mu \psi_{xt} = 0$&$z = k x - c t$ &$- \frac{k (1 + \lambda \mu)^2 + c \lambda (2 + \lambda \mu)}{c^2 \mu (c + k \mu)} f^2 $& +\\
Wadati, Segur,  & &$- \frac{2 C_1}{c (c + k \mu)}$, $C_1$\ const.  & \\
Ablowitz \cite{WaSeAb}& &   &\\\hline

\caption*{If condition (\ref{Khare.f6}) can be fulfilled (e. g., by
choosing an appropriate constant of integration) start solutions for superposition can be obtained by Eqs. (\ref{Simple.f4}), (\ref{Simple.f10})
(marked by a "+", otherwise by a "-").}
\end{longtable}
}


\section*{Acknowledgements}
One of us (J. N.) gratefully acknowledges support by the German Science Foundation (DFG)
(Graduate College 695 "Nonlinearities of optical materials").

\appendix

\section{Superposition solutions of the nonlinear cubic-quintic Schrödinger equation (NLCQSE)}\label{AppA}

Following the lines described in Secs. \ref{Intro}, \ref{StartSol} the NLCQSE

\begin{equation}\label{Khare.f8}
 i \psi_{t} + \psi_{xx} - (q_1 |\psi|^2 + q_2 |\psi|^4)
\psi = 0, \quad
\end{equation}

($q_1$, $q_2$ real constants) can be solved by applying the transformation

\begin{equation}\label{Khare.f9}
\psi(x,t)=f(z)\exp[i(\lambda t + r(z))], z = x - ct.
\end{equation}

Separating real and imaginary parts, we obtain

\begin{equation}\label{Khare.f11}
q_1 f(z)^3 + q_2 f(z)^5 - f''(z) + f(z) (\lambda - c r'(z) +
r'(z)^2) = 0, \quad
\end{equation}

\begin{equation}\label{Khare.f12}
 f'(z) (c - 2 r'(z)) - f(z) r''(z) = 0,
\end{equation}

where the prime denotes differentiation with respect to $z$.

Equation (\ref{Khare.f12}) can be integrated to yield

\begin{equation}\label{Khare.f13}
  r'(z) = \frac{c}{2} + \frac{C_1}{f(z)^2}, \quad
\end{equation}

with $C_{1}$ a constant of integration.

Inserting $r'(z)$ into Eq. (\ref{Khare.f11}) and integrating the resulting expression leads to an equation where $h = f^2$ can be introduced.
Thus, we find a basic equation $R(h)$ (cf. Eq. (\ref{Out.f1}), $f \rightarrow h$) with the following coefficients:

\begin{equation}\label{NLSE.f1}
\alpha = \frac{4}{3}\: q_2,\: \beta = \frac{1}{2}\: q_1,\: \gamma = \frac{4 \lambda - c^2}{6},\: \delta = 2\: C_2,
\epsilon = - 4 \:C_1^2,
\end{equation}

where $C_2$ is a constant of integration.\\
If $q_2 = 0$ and $C_1 = 0$ all physical solutions suitable for superposition are represented by Eqs. (\ref{Simple.f4})
($f \rightarrow h$). The superposition solutions for $p = 3$  are given by (cf. Eqs. (\ref{Khare.f2}), (\ref{Khare.f9}))

\begin{eqnarray}\label{Sup.f29_1}
\widetilde{\psi}(x,t) &= a \sum_{i = 1}^3 \textrm{cn}\left[  \mu
(x - v_3 t) + \frac{4 (i - 1) K(m)}{3}, m\right]
\exp\left\{i\left[\lambda t + (x - v_3 t) \frac{v_3}{2}\right]\right\},\nonumber\\
v_3^2 &= 4 (\lambda - \mu^2 (2  m  X(3,m) + (2 m -1))),
\end{eqnarray}

\begin{eqnarray}\label{Sup.f4_1}
\widetilde{\psi}(x,t) &= a \sum_{i = 1}^3 \textrm{dn}\left[ \mu
(x - v_3 t) + \frac{2 (i - 1) K(m)}{3}, m\right]
\exp\left\{i\left[\lambda t + (x - v_3 t)\frac{v_3}{2}\right]\right\},\nonumber\\
v_3^2 &= 4 (\lambda + \mu^2 (m - 2) - a W(3,m)),
\end{eqnarray}

\begin{eqnarray}\label{Sup.f11_1}
\widetilde{\psi}(x,t) &= a \sum_{i = 1}^3 \textrm{sn}\left[ \mu
(x - v_3 t) + \frac{4 (i - 1) K(m)}{3}, m\right]
\exp\left\{i\left[\lambda t + (x - v_3 t)
\frac{v_3}{2}\right]\right\},\nonumber\\
v_3^2 &= 4 (\lambda +  \mu^2 (m + 1) + 2 m a \mu^2 V(3,m)).
\end{eqnarray}

To evaluate the speed $v_3$ we have used in Ref. \cite{38} the Eqs. (8), (70), (72),
Eqs. (8), (66), (68) and Eqs. (8), (57), (59), respectively.\\
It should be mentioned that the start solutions (\ref{Simple.f4}) suitable for superposition are consistent with those of Cooper, Khare
and Sukhatme \cite{38}. Nevertheless, the speed $v_3$ according to Eqs. (\ref{Sup.f29_1}), (\ref{Sup.f4_1}), (\ref{Sup.f11_1}) is not identical
with $v_3$ according to Eqs. (33), (28), (45) in Ref. \cite{38}. Thus, the superposition solutions are not determined uniquely.
Different identities between Jacobian elliptic functions used lead to (in general) different superposition solutions.
Applying the procedure outlined in Sec. \ref{StartSol} if $q_2 \neq 0\: (\alpha  \neq 0)$, $\beta = \delta = 0$,
$\epsilon$ arbitrary, PDC implies either $q_2 = 0$ ($\alpha = 0$) or $C_1^2 = 0$ ($\epsilon = 0$).
The choice $q_2  = 0$ (in addition to $q_1 = 0$ ($\beta = 0$)) is not of interest, because it leads to a
linear equation (\ref{Khare.f8}). For $C_1^2 = 0$ we obtain solitary traveling-waves.
Thus, since $\psi(x,t)$ is not periodic,
superposition solutions are not possible in this case.


\begin{thebibliography}{99}
\bibitem{9a} We assumed $\triangle > 0$. If $\triangle < 0$,
substitution of $\wp(z)$ according to Ref. \cite{9}, 18.9.11 does
not lead to an expression of form (\ref{Khare.f1}).
Furthermore, as will be seen below, we obtain a polynomial $R(f)$ or $R(h)$, $h = f^2$, of third degree for generating
new solutions by linear superposition. But the PDC
is not satisfied for a third-degree polynomial with $\triangle < 0$, because two of the three
roots are complex conjugate.
\bibitem{9b} Since we are interested in physical periodic solutions we can always assume that a simple root exists.
\bibitem{9aa} If $\beta = 0$ holds $\gamma$ must be negative
otherwise Eq. (\ref{Khare.f6}) and PDC are not fulfilled. For $f_0$ to
be a simple root $\delta^2 - \frac{3}{2} \epsilon \gamma$ must be
positive. If $\beta \neq 0$ the discriminant $\triangle$ does not
vanish, so that $f_0 = \frac{2 e_3 - \gamma}{2 \beta}$ is a simple
root of $R(f)$.
\bibitem{53} If parameters according to Eqs. (\ref{KP.f5}) are inserted into Eq. (\ref{Simple.f10}) one obtains
$B$, $\mu$, $m$ so that $6 B c m \mu^3 - m \mu b B^3$ in Eq. (\ref{KP.f17}) vanishes identically.
\bibitem{Sug1}  An ansatz
$\widetilde{\psi} = \textrm{cn}$ or $\widetilde{\psi} = \textrm{sn}$ (cf. eq. (\ref{KP.f13}))
leads to equations which have the form (\ref{KP.f14}). Because
there is no relation $\textrm{c}_1 \textrm{c}_2 = \textrm{const.}$ and $\textrm{s}_1 \textrm{s}_2 = \textrm{const.}$,
respectively, there is no possibility to replace the appearing sums $\sum
\textrm{c}_i \left(\sum \textrm{d}_i \textrm{s}_i\right)^2$, $\left(\sum \textrm{c}_i\right)^2
\frac{d}{dx} \sum \textrm{d}_i \textrm{s}_i$ and $\sum \textrm{s}_i \left(\sum \textrm{c}_i
\textrm{d}_i\right)^2$, $\left(\sum \textrm{s}_i\right)^2 \frac{d}{dx} \sum \textrm{c}_i
\textrm{d}_i$, respectively. Up to our knowledge there is no appropriate relation involving
Jacobian elliptic functions that would simplify the equations similar to (\ref{KP.f14}), so that the
speed $v_2$ for which $\widetilde{\psi}$ is a solution of eq. (\ref{KP.f1}) cannot be determined.
\bibitem{Burgers} As outlined above start solutions for linear superposition can be obtained if $\alpha = 0$
or $\beta = \delta = 0$. In the case of the Burgers equation these conditions lead to $k \rightarrow \infty$ or $c = 0$ for a
traveling wave ansatz $\psi = f(k x - c t)$.
\bibitem{KdVB} As outlined above start solutions for linear superposition can be obtained if $\alpha = 0$
or $\beta = \delta = 0$. In the KdV-Burgers equation the only parameter
which can be varied in the basic equation is $k$. If $k \rightarrow \infty$ then $\alpha \rightarrow 0$, but also
$(f_z)^2 \rightarrow 0$ so that $f \equiv \textrm{const.}$.
\end{thebibliography}
\end{document}